\documentclass[rmp,aps,nofootinbib,preprint,draft,showpacs,amsmath,amssymb]{revtex4}
\usepackage{epsfig}
\usepackage{epsf}
\newcommand{\bea}{\begin{eqnarray}}
\newcommand{\eea}{\end{eqnarray}}

\def\myfigure#1#2{{\leftskip=0.000753\textwidth
\rightskip\leftskip\small \begin{figure}\baselineskip=14pt plus 2pt
minus 1pt \centerline{#1}\nobreak\smallskip\nobreak #2\end{figure}}}

\begin{document}
\title{Scale invariance in coarsening of binary and  ternary fluids }
\author{K.C. Lakshmi \footnote{Email:lakshmi@physics.iitm.ac.in} and 
 P.B. Sunil Kumar\footnote{Email:sunil@physics.iitm.ac.in}}
\affiliation{Department of Physics,\\
Indian Institute of Technology Madras,\\
Chennai 600 036,\\
India}                                   
\date{\today}

\begin{abstract}    
Phase separation in  binary and ternary
fluids is studied using  a two dimensional Lattice Gas Automata. 
The lengths, given by the the first zero crossing point of 
the correlation function and the total interface length is shown to exhibit 
 power law dependence on time.  
In binary mixtures, our data clearly indicate 
the existence of a regime having more than one length scale
where the coarsening process proceeds through the rupture and 
reassociation of domains. In ternary fluids; in the case of symmetric 
mixtures there exists a regime with a single length scale having dynamic
 exponent $1/2$, while in asymmetric mixtures our data establish the
 break down of scale invariance.

\end{abstract}    
\maketitle

\section{Introduction}
A mixture of fluids phase separate into domains when quenched below its 
critical temperature. Since many systems of scientific and technological 
interest are multi component mixtures, the phase behavior of
fluid mixtures are of current interest. The equilibrium state of 
incompatible fluids is one in which the pure phases are separated by  
a single connected interface. However, in the 
thermodynamic limit, starting from a mixed phase this 
equilibrium is never reached. In view of this, the kinetics of the
phase separation process gains importance.

In the case of phase separating  binary fluids, experiments~\cite{wong-knobler81,chou-goldburg81,perrot-guenoun94,walheim99}
and numerical simulations~\cite{velasco-toxvaerd93,leptoukh-strickland95,
valls-farrel89} have clearly established the importance of hydrodynamics 
in the determination of late time domain growth laws.
In spite of all  efforts,  a complete theoretical understanding of this 
highly nonlinear phenomena remains unsolved~\cite{bray94}.  This is mainly 
due to the fact that studying kinetics of  liquid phase separation involves 
the solution of mutually coupled equations; the Navier-Stokes equation
for the flow and the  equation of continuity  for the order parameter.

Early experimental and theoretical studies of these binary fluid systems, 
assume that
 there exists, at late  times, a dynamical scaling regime exhibiting  similar 
behavior under an appropriate rescaling of time and length scales.
Dynamical scaling is characterized by the single time-dependent length
scale $R(t)$.  The domain growth follows a simple and generic algebraic 
form, $ R(t) \propto  t^{\alpha}$ ,where $\alpha$ represents the exponent 
characteristic of the universality class to which the system belongs. 
Scaling and dimensional analysis by Siggia ~\cite{siggia73} ,
Furukawa ~\cite{furukawa94}, San Miguel~\cite{sanmiguel-grant85}  and more 
recently by Bray~\cite{bray94} addresses this question of the growth exponent
taking this length scale $R(t)$ to be the average size of the 
ordering domains.
In two dimensions, in the case of  minority B phase separating from a A rich mixture , 
the domains coarsen via evaporation-condensation
process leading to the exponent $\alpha=1/3$. The late time 
growth is governed by  droplet coalescence  leading to the exponent 
$\alpha=1/2$.  In the case of symmetric binary mixtures, thermal 
fluctuations can  drive the initial $\alpha=1/3$ regime to a 
$\alpha=1/2$ regime  ~\cite{sanmiguel-grant85}.  Most  of the studies, 
however show a cross over from the $\alpha = 1/3$ to an inertial 
$\alpha=2/3$ regime predicted by  Furukawa~\cite{furukawa94}.

More recently many researchers have pointed out  the absence of scaling  in two 
dimensional  phase separating fluids. It is shown that competition between 
diffusive and hydrodynamic growth leads to breakdown of scale-invariance in 
symmetric binary fluids ~\cite{wagner-yeomans98}. In the viscous hydrodynamic 
regime, scaling is observed only 
in the case of coarsening through coalescence.  Even after starting from a 
droplet state one can enter a bicontinuous state by coalescence induced 
coalescence mechanism ~\cite{tanaka94} leading to a breakdown of scaling.
Alternatively starting from a droplet state scaling 
is maintained, in symmetric binary mixtures, if the droplet morphology 
is self sustaining ~\cite{wagner-cates01}.  
In the inertial hydrodynamics regime ,full scaling is recovered 
~\cite{furukawa00,wagner-cates01}. Excellent scaling 
is observed also in the cross over regime from the viscous hydrodynamic to the 
inertial hydrodynamic regime ~\cite{pagonabarraga01}.

Variety of techniques have been used to simulate growth kinetics in 
the binary immiscible fluids.  Direct simulations of model H have been used 
to study the domain growth and scaling using variously defined length 
scales ~\cite{furukawa00,bray94}.  Lattice Boltzmann simulations have been 
particularly useful in exploring the late time hydrodynamic regime
~\cite{yeoman1-96,chen93,yeoman3-98}. One problem with this technique is 
that it does not include thermal fluctuations. On the other hand thermal 
fluctuations are 
inherent in Lattice Gas models. Rothman and Keller proposed a Lattice Gas 
Model (RK model) for the binary immiscible fluids~\cite{rothman-keller88}.
This model has been used for simulating 2-D binary fluid  phase separation at 
different overall fluid densities ~\cite{bastea-lebowitz95,coveney-emerton97}.

Though most of the 
techniques mentioned in the previous paragraph have been extended to study 
 a mixture of two fluids and a surfactant ~\cite{coveney-emerton97,kawakatsu93, Gonnella98}, 
our understanding of phase separation wherein all three
are fluid components is rather limited.
Molecular Dynamics simulations by Laradji 
{\em et.al} ~\cite{laradji-mouritsen94} 
seem to indicate that 
hydrodynamic flow is not likely to control the separation process in 
ternary fluid system. This study indicates that the ternary system  
at late times reaches a dynamical scaling regime during which the 
domains show a growth law $R(t) \propto t^{1/3}$, in agreement with the 
classical theory of Lifshitz and Slyozov.  
Gunstensen and Rothman extended the  Lattice Gas Model (GR Model) 
to study ternary 
immiscible fluids (LGTF)~\cite{gunstensen-rothman90} but do not make 
any comments about the dynamical exponents. In this model the total energy 
function is the sum of the work done by the ``color flux" ${\vec f_i}$ of each
component against its  ``color field" ${\vec Q_i}$ and is given by 
$W=\sum \sigma_i {\vec f_i} \cdot {\vec Q_i}$.
More recently, a level set method  was proposed for the study of 
phase separation in fluid mixtures~\cite{smith00}. In this method 
one assumes convective terms to dominate the coarsening; local volume fraction   function $\phi_i(x,t)$  of the components is then coupled to the local 
velocity field ${\bf v}$ through the kinetic equation, 

\begin{equation}
    \frac{\partial\phi}{\partial t} \,=\,- {\bf v}.\nabla \phi .
\end{equation}

The fluid velocity ${\bf v}$ satisfies  Navier-Stokes 
equation 
 
\begin{equation}
   \frac{\partial{\bf v}}{\partial t} + ({\bf v}.\nabla) {\bf v}
  = \nu \nabla^{2} {\bf v} - \frac{\nabla P}{\rho} -\frac{F}{\rho},
\end{equation}.

Here $P$ is the pressure ,$\nu$ is the kinematic viscosity and $\rho$ is 
the density. The interfacial energy between the domains enters the equations
only through the external force $F$; 

\begin{equation}
\bf{F}\,=\,\sum_{i=A,B,C} \sigma_i' \delta(\phi_i) \kappa(\phi_i) \hat{\bf n},
\end{equation}

\noindent
where $\kappa(\phi_i)$ is the curvature of the $\phi_i$ domain interface, $\sigma$ is the 
surface tension defined such that for the $AB$ interface the surface tension is 
$\sigma_A^i+\sigma_B^i$.\\

\noindent
The zero contour of the function $\phi_i$ specifies the interface of the domains. 
Immiscible three component mixtures with majority components A and B having the same volume 
fraction and a minority C phase was studied using this method ~\cite{smith00}. They find 
a power law dependence of interface length  with dynamic exponents in the  
range of $0.5 \sim 0.6 $.
   
In this paper we discuss  Lattice gas simulations of
the binary and ternary mixtures phase separating  in two dimensions.
We calculate cluster size distribution,total interface length  and the 
density density correlation functions for each of the components. We 
see that, in some cases, the interface length and the first zero of the 
correlation function; though shows power law dependence on time, do not 
have the same dynamic exponents, thus violating scaling.

In binary mixtures large difference in the parameters $\sigma_i$ of
the two components can lead to break down of scaling with the domains 
coarsening through rupture and coalescence.
In the symmetric ternary mixtures with equal $\sigma_i$ 
and equal volume fractions,
we see  the average size of the ordering domains, $R(t) \propto t^{1/3}$ 
in agreement with the  results reported earlier. 
At late times, we observe a crossover to a  $t^{1/2}$ regime which 
is not reported in previous simulations. As we reduce the volume fraction of 
one of the components this exponent changes to $2/3$, consistent with 
the inertial regime in two component mixtures.  
In the asymmetric case; wherein one of the component is 
miscible in the other two, we observe a $R(t) \propto t^{1/2}$  
and a cross over to $t^{2/3}$ as  the volume fraction of the 
solute is reduced to zero.

By choosing a substrate which has minimum interaction with the fluid one 
should be able to conduct experiments on fluid separation problem in two 
dimensions. Such quasi two dimensional geometries have been used for 
studying three fluid phase separation ~\cite{walheim99}.

The paper is organized as follows:
In section I,we introduce the problem and summarize previous results. 
A detailed account of the 
Lattice Gas Automaton model and the algorithm for computing the cluster 
size and interface length are given in section II. Next, we discuss the 
results of our simulations on the binary  fluids. Section IV deals with  
the symmetric and asymmetric ternary fluids.
In section V,we conclude the paper with a  summary of the  results.

\section{The model}

  Lattice Gas Automaton (LGA) is an alternative numerical description of fluid  flow
  dynamics. This model approximates reality by constraining motions and 
collisions of
 fluid particles to a lattice, where each particle represents a finite mass of
 fluid. 
  Lattice-gas automaton models represent, in the continuum, 
 the incompressible Navier-Stokes equation correctly and are extensively used 
in simulations of fluids.

In this paper we discuss simulations using the FHP model for fluids. 
Much has been written about the FHP model which was first introduced in 1986 
by Frisch, Hasslacher and Pomeau~\cite{chopard-droz98}. The fully discrete 
microscopic dynamics of the FHP model maps into the macroscopic behavior of 
hydrodynamics.  A modified version of the FHP model for immiscible fluid was introduced by 
Rothman and Keller~\cite{rothman-keller88}. 
  In this model, which is defined on a hexagonal lattice,  particles  at 
the lattice sites are allowed to take 7 
possible velocities, $\vec{c_{i}}, i \in \{0,1,2,3,4,5,6\}$ .
 Boolean variables indicate the occupation number at a particular site at a
 given time.  The dynamics is such that no more than one particle enters 
the same site at the same time with the same velocity; the exclusion principle. 
The particles undergo collision that conserves, 
 at every site, the total number of particles,their total momentum and the 
total kinetic energy  after each time step.

 Multi phase flows in which different species of fluids coexist, move 
and interact are an important domain of application of the lattice gas approach.
Models for miscible, immiscible and reactive  flows have been proposed.  Interaction
between the species are the key ingredient of multi-phase flows and depending
on the nature of the physical processes, different interactions will be
considered.

The Lattice gas approach has the important advantage that the interface between
the different fluids appear naturally as a consequence of the way the fluids 
are modeled in terms of particles. 
In this paper, we represent the different 
phases by assigning the particles a ``color" .
Now the state of any vertex  at $\vec{r}$ is specified by a total density  vector 
$\vec{\phi}(\vec{r})$ and 
the color vectors $\vec{q_i}(\vec{r})$ with Boolean variables $\phi_{j}$ and $q_{ij}$ as 
the components of this vector.
In the case of ternary mixtures,  $i$ can take 3 values +1 , -1 and 0 .  
This is illustrated clearly in figure 1. We define the average density as 
the ratio of total number of particles  to the maximum number of particles 
possible. This is given by $d=\sum_{i=0,j=1}^{i=6,j=N}\phi_i(r_j)/(7 N)$, where
$N$ is the total number of sites in the lattice.

\myfigure{\epsfysize1.5in\epsfbox{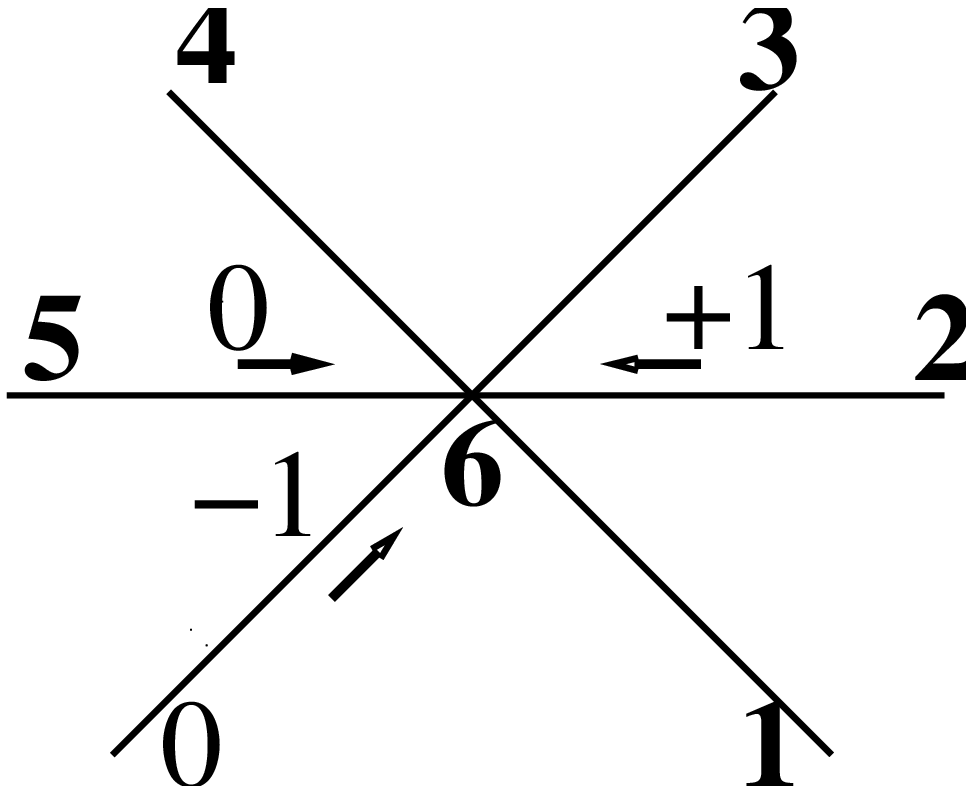}}{\vskip-.0in
Fig.\ 1~~ The {\em state} of a lattice point is given by the vectors $\vec{\phi}=\{1,0,1,0,0,1,0\}$ , 
$\vec{q_1}=\{0,0,1,0,0,0,0\}$, $\vec{q_{-1}}=\{1,0,0,0,0,0,0\}$ and $\vec{q_0}=\{0,0,0,0,0,1,0\}$}

Drawing analogies from electrodynamics, we define a ``color flux" and
 ``color field"at every lattice point  for each type of particles. 
In the case of  ternary mixture the color flux at a lattice site is given by

\begin{equation}
\vec{Q_{1}}(\vec r)=\sum_{j=0}^{j=5}\vec{c_{j}}q_{1j},
\end{equation}

\begin{equation}
\vec{Q_{-1}}(\vec r)=\sum_{j=0}^{j=5}\vec{c_{j}}q_{-1j},
\end{equation}

\begin{equation}
\vec{Q_{0}}(\vec r)=\sum_{j=0}^{j=5}\vec{c_{j}}q_{0j}.
\end{equation}
And the local color gradients or fields are defined to be

\begin{equation}
\vec{f_{1}}(r)=\sum_{j=0}^{j=5}\vec{c_{j}}\sum_{k=0}^{k=6}q_{1k}(\vec r+c_j)
\end{equation}

\begin{equation}
\vec{f_{-1}}(r)=\sum_{j=0}^{j=5}\vec{c_{j}}\sum_{k=0}^{k=6}q_{-1k}(\vec r+c_{j})
\end{equation}

\begin{equation}
\vec{f_{0}}(r)=\sum_{j=0}^{j=5}\vec{c_{j}}\sum_{k=0}^{k=6}q_{0k}(\vec r+c_{j})
\end{equation}

 Following the Gunstensen and Rothman (GR model), we write the work done by the flux 
against the field to be:

\begin{equation}
W=-(\sigma_1 \vec{f_{1}}(r) . \vec{Q_{1}}(r) +\sigma_{-1} \vec{f_{-1}}(r) . \vec{Q_{-1}}(r)
   +\sigma_0 \vec{f_{0}}(r) . \vec{Q_{0}}(r) \label{three-energy})
\end{equation}

where $\sigma_i$  are the parameters which determine the surface tension between
 different phase boundaries. The three phases are labeled by $i=\pm1,0$.

The simulation proceeds  in two steps, Collision and Translation.
During the collision  process, particles at a given lattice site can exchange 
their velocity and color.
The collision rules  are such that the work performed by the flux against the field is minimized , 
subject to the constraints of the exclusion principle as well as the principles
 of mass ,momentum and  particle conservation. 
There can be a number of  configurations which satisfy the above constraints, and in principle one 
should pick up that configuration which minimizes energy from this full set of possible configurations.
In practice only a ``limited" number  of configurations are considered. In our simulations we take into account 
the following possibilities.

If a site has more than three particles or if their total momentum is non zero,
then we pick up all possible pairs and an attempt is made to exchange their 
colors. 
The  exchange is accepted only if it lowers the energy.

\myfigure{\epsfysize1.7in\epsfbox{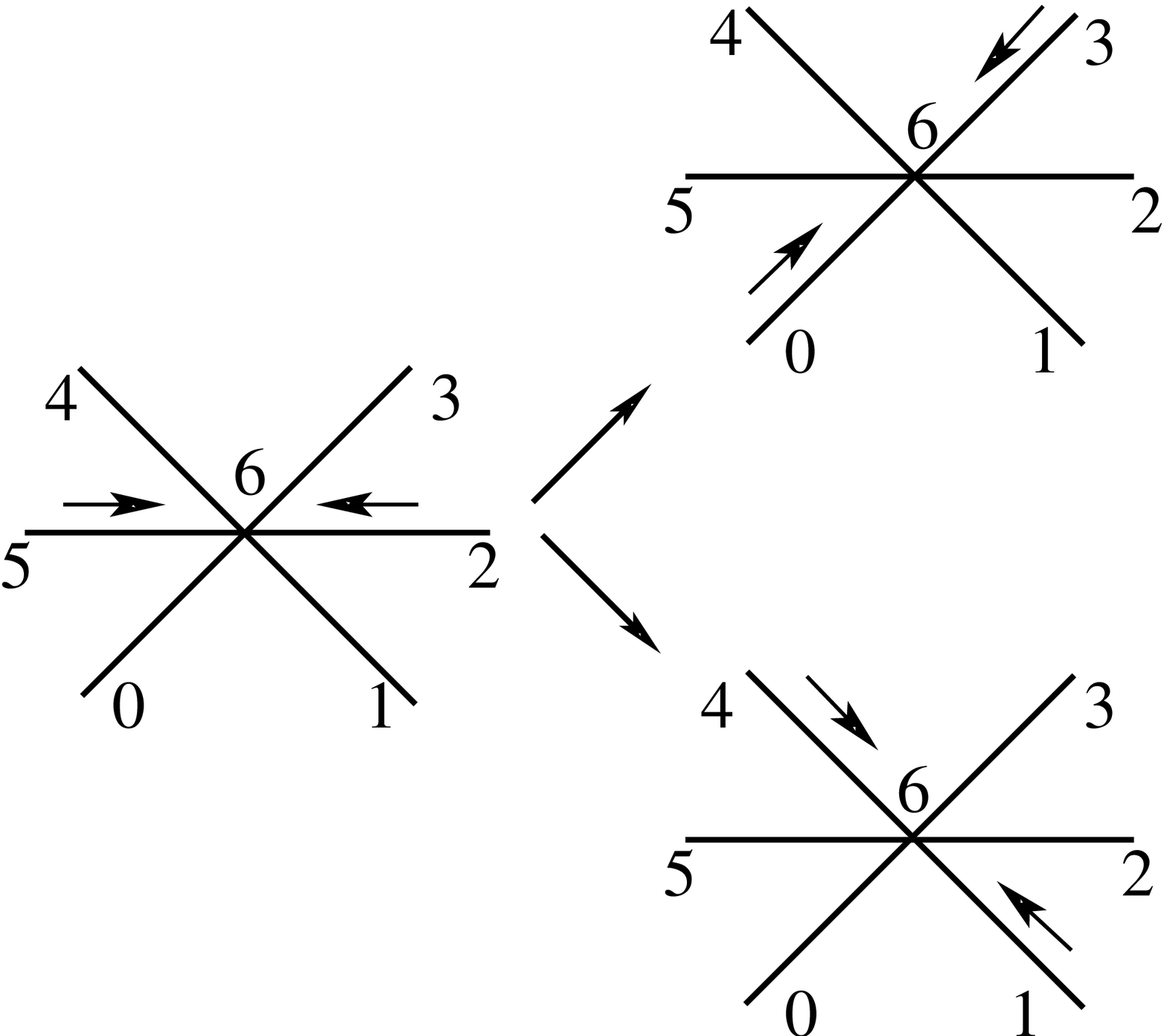}}{\vskip-.0in
Fig. \ 2~~ Possible configurations in a two particle collision }

When 2 particles occupy the same site with opposite 
velocities there are two cases to consider. In one both the particles are of the 
same color. Then a rotation of the configuration by $0,60$ or $120$ degrees is 
attempted as shown in  figure 2. 
The other possibility is that the two are of different colors. Then we need 
to check the possibilities for $0, \pm 60, \pm 120$ and $ 180$ degree rotation.

        When three particles collide at an angle of 120 degrees there are 
three possibilities. When all the particles have the same color,
they could just reverse their directions or  retain their velocities. When 
the colors are not the same, similar to  two particle collisions, the 
configuration could now be rotated by  $0, \pm 60, \pm 120$ and $ 180$ .
In all the above cases the lowest energy configuration is chosen as the final 
state.

There is one special case of three particle collision in which two particles 
have opposite velocities and the third one is a  zero velocity particle. In this case, 
apart from rotating the configuration  an exchange of color between pairs 
is also attempted. 

   Next step in the simulation is the translation in which all particles 
are moved in the direction of their velocities. This collision and translation
process completes one time step. 

To understand the various competing terms in this model we first look 
at a two component fluid mixture and
introduce the variables $q_{k}=q_{1k} - q_{-1k}$, 
$\phi_{k}= q_{1k} + q_{-1k}$, in the direction $k$. The two components are now designated by 
$q_k=\pm 1 $. Substituting these variables in the equations for 
$\vec{f_1},\vec{f_{-1}}$ and $\vec{Q_1}, \vec{Q_{-1}}$, we can obtain from 
equation~\ref{three-energy} the work done 
to be 
\begin{eqnarray}
		W& =&-( \sigma_{1} f_{1} Q_{1} +\sigma_{-1} f_{-1} Q_{-1})\nonumber\\
		 &=&-( \sum_k c_k q_k \left((\sigma_1+\sigma_{-1})\sum_{i}
		 c_i \sum_l q_l(r+c_i) + (\sigma_1 -\sigma_{-1})
			\sum_i c_i \sum_l \phi_{l}(r+c_i)\right) \nonumber \\
		 && +\sum_k c_k \phi_k \left((\sigma_1+\sigma_{-1})\sum_{i}
		 c_i \sum_l \phi_l(r+c_i) - (\sigma_1 -\sigma_{-1})
			\sum_i c_i \sum_l q_{l}(r+c_i)\right)) \label{three-energy-1}
\end{eqnarray}

Momentum conservation implies that the last two terms in the above 
equation do not contribute to the dynamics. The first term  decides the 
energy of the interface between the two phases. The second term favors 
movement of particles toward higher or lower densities depending on 
their color. For example when $\sigma_1-\sigma_{-1} > 0$ this terms imply 
higher diffusivity of particles with $q=-1$. Thus increasing 
$|\sigma_1-\sigma_{-1}|$ is like increasing the ``temperature".

The earlier lattice gas model for two component fluids  
by Rothman and Keller (RK model) used
$\sigma_1=\sigma_{-1}=\sigma$. In 
this model the work done is given by
\begin{equation}
	W = -\sigma \vec{f}\cdot  \vec{Q} \label{two-energy},
\end{equation}

where $\vec{Q}(\vec r)=\sum_{j=0}^{j=5}\vec{c_{j}}q_{j}$ and 
$\vec{f}(r)=\sum_{j=0}^{j=5}\vec{c_{j}}\sum_{k=0}^{k=6}q_{k}(\vec r+c_j)$.
We can then  introduce thermal fluctuation by defining 
an inverse temperature like parameter $\beta$ such that the new configuration 
at every time step is accepted with a probability proportional 
to $\exp{-\beta (W_{old}-W_{new})}$ ~\cite{coveney-emerton97}. Unless 
specified otherwise the results discussed in this paper are obtained 
using the GR model.

\subsection{Correlation functions and domain distribution functions}

We define the  pair correlation functions
\begin{equation}
 C^q_{ij}(\vec{r},t) =\left<\sum_k q_{i,k}(\vec{x},t)\sum_kq_{j,k}(\vec{x}+\vec{r},t)^{}\right>- \left<\sum_k q_{i,k}(\vec{x})\right>\left<\sum_k q_{j,k} (\vec{x})\right>,
\end{equation}
\begin{equation}
 C^h_{ij}(\vec{r},t) =\left<h_i(\vec{x},t)h_j(\vec{x}+\vec{r},t)^{}\right>- \left<h_i(\vec{x})\right>\left<h_j(\vec{x})\right>,
\end{equation}

Here the subscripts $i,j$ represent the color of the particles with  $i,j=\pm 1,0$.
Field values $h_i$ determines the {\em phase} at every 
lattice points. A lattice point belongs to the $i^{th}$ phase if 
$h_i$ has a value higher than the other two. This means that most of the 
particles there belong to the $i^{th}$ color. We calculate the length
$R_{ij}(t)$ as the first zero of the correlation function $C^h_{ij}$.

\myfigure{\epsfysize3.4in\epsfbox{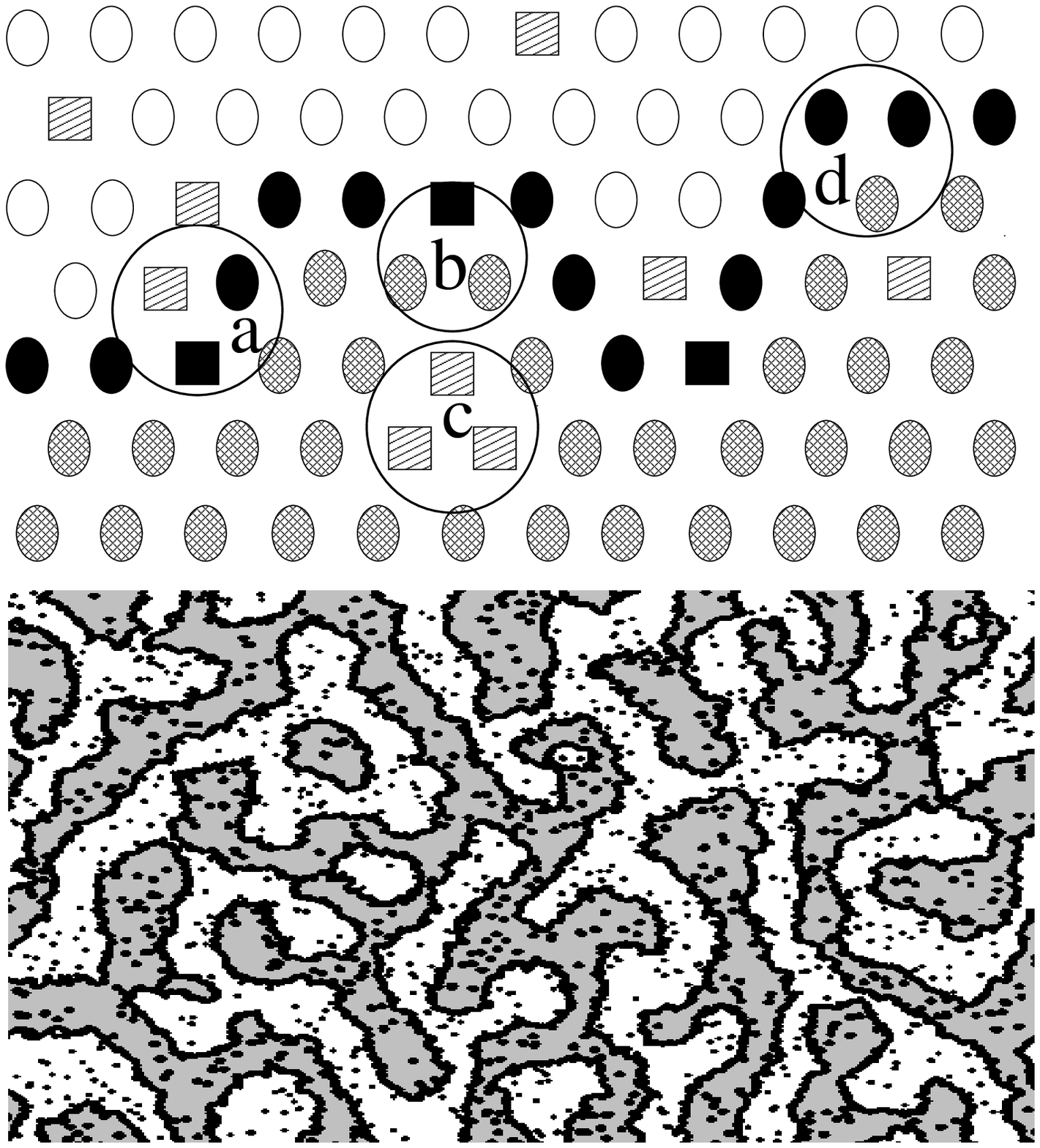}}{\vskip-.0in
Fig. \ 3~~ (top) Rules for determining the border. Square represent C phase,
Open circles represent B phase and the hashed circles represent A phase. The
 dark squares and circles are the border sites between the A-C phase and the 
B-C phase. (bottom) Snap shot from the simulation of ternary fluid with one 
component miscible in the other two. The dark regions are  the border 
between th A-C (white) and B-C (gray) domains. }

For the immiscible case wherein all three phases have the same surface tension,
to calculate the border length and cluster size of the domains of a particular 
phase, say A, we use the following algorithm. 

All lattice sites are picked 
sequentially. If a particular site belongs to the phase A and is not assigned a 
cluster number we give it a new number. All the neighbors of this site are
 then checked. If the neighbor belongs to the phase A, three possibilities 
arise. In one, the site does not have an  assigned cluster number. We will then 
give this neighbor the same cluster number as the present site. In the 
second case, the neighbor has a number which is the same as the site . Here 
we do not have to do anything more. In the third case the neighbor has a
 number which is different from that of the present site.
That is, the neighbor belongs to a different cluster. In that case, the 
clusters are merged by assigning  smaller of  the two numbers  to all sites 
in the two clusters.  A lattice site which belong to a particular 
phase is a border site if one of its neighbors is of a different phase.

In the case, wherein one of the phases is equally miscible in the other two; 
for example let us consider C miscible in A and B, we have two phases, the 
A-C phase and the B-C phase, see figure 3. To label the domains we then 
follow the same procedure as before. Any site belongs to the A-C phase  
if it is an A site or a  C site whose  neighbors are all A or C. For example 
region c in figure 3 belongs to the A-C phase. The border 
of the A-C domain could be a A or C site. If it is an A site, it should have 
at least one B neighbor (region d in figure 3) or  
a C neighbor  with the next nearest neighbor as B (region a in figure 3).
If the border site belongs to the C phase then it should either have 
a neighbor which is a B site (region b in figure 3) or a C neighbor  with the next nearest neighbor as B (region a in figure 3).

  In the simulations, to prepare the initial configuration 
a particle is placed at a  randomly chosen site.  The velocity of the particle
is chosen randomly from the seven possible directions such that the total 
momentum summed over all particles is zero. The color of the particle is 
again assigned randomly depending on the volume fraction of each component.
Once the initial configuration is prepared the ensuing dynamics  conserved
the total number of particles, the net momentum, the total kinetic energy and the volume fraction of each components.

	Simulations are performed on a triangular lattice with 360000 lattice 
points with periodic bondary conditions. The results are averaged over 
10 initial configurations. The dynamic 
exponents are obtained by a power law fit over one decade in time. 	

\section{ Two component fluids}

In this section we reexamine the well studied case of  binary fluid
 coarsening following a
 critical quench. The fluids are labeled by $i=\pm 1$. Let us first look at 
the case $\sigma_1=\sigma_{-1}=1$. The time dependence 
of average domain size $R(t)$ which scales the same way as $R_{ij}$ is shown 
in figure 4.  At early time we see Lifshitz-Slyozov-Wagner(LSW)  $t^{1/3}$ 
~\cite{bray94} growth. At late times the growth of domains is 
controlled by the inertial hydrodynamics and we have the well known 
$t^{2/3}$ growth law ~\cite{bray94}.

\myfigure{\epsfysize1.8in\epsfbox{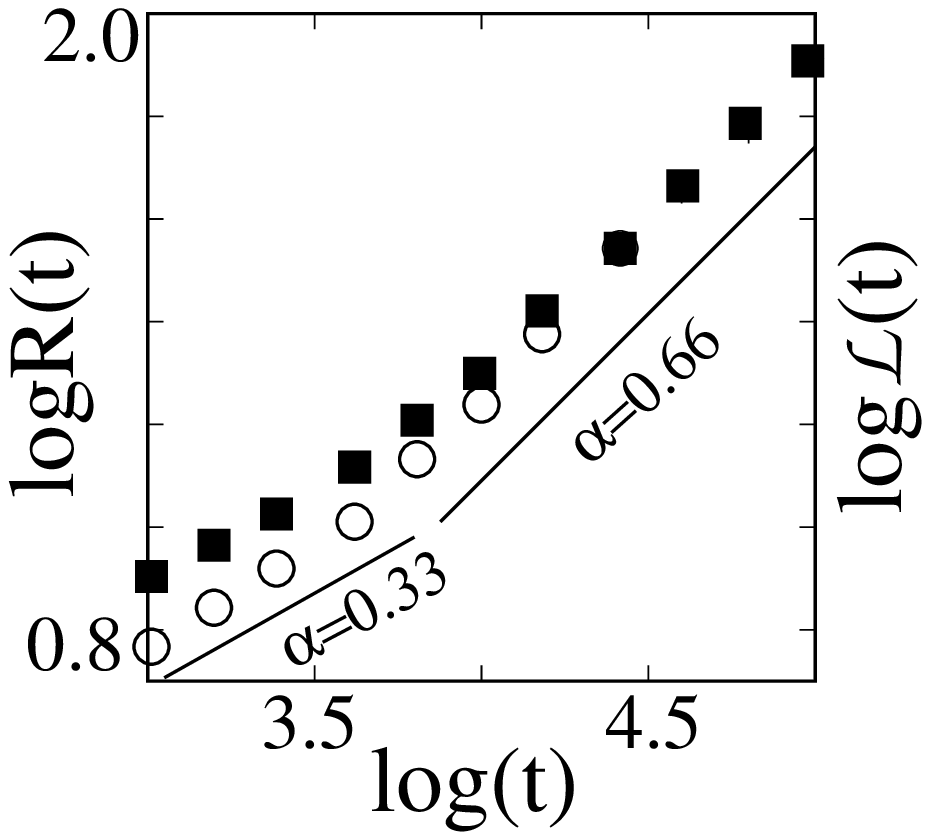}}{\vskip-.0in
Fig.\ 4~~ The first zero of the correlation function $R$ 
(open circles) and the total 
border length ${\cal L}$ (filled squares) for a binary fluid, with $\sigma_1=\sigma_{-1}=1$, as a function 
of time. The average density of the fluids is $d=0.55$. 
The curves are shifted to make the comparison easy. Continuous 
lines with  slope $2/3$  and $1/3$ are given as guide to the eye   }

To check for scaling  we compare the length $R$ given by the first zero of the 
correlation function with  the  total border length 
${\cal L}(t)$ . As can be seen from figure 4, at late time, in the 
inertial hydrodynamic regime, 
the length ${\cal L}$ also exhibits  power law dependence with a dynamic 
exponent $2/3$.  This agrees with earlier studies on 
binary fluid phase separation which predict scaling to hold  
in the inertial 
regime~\cite{wagner-cates01,furukawa00}. 

\myfigure{\epsfysize1.8in\epsfbox{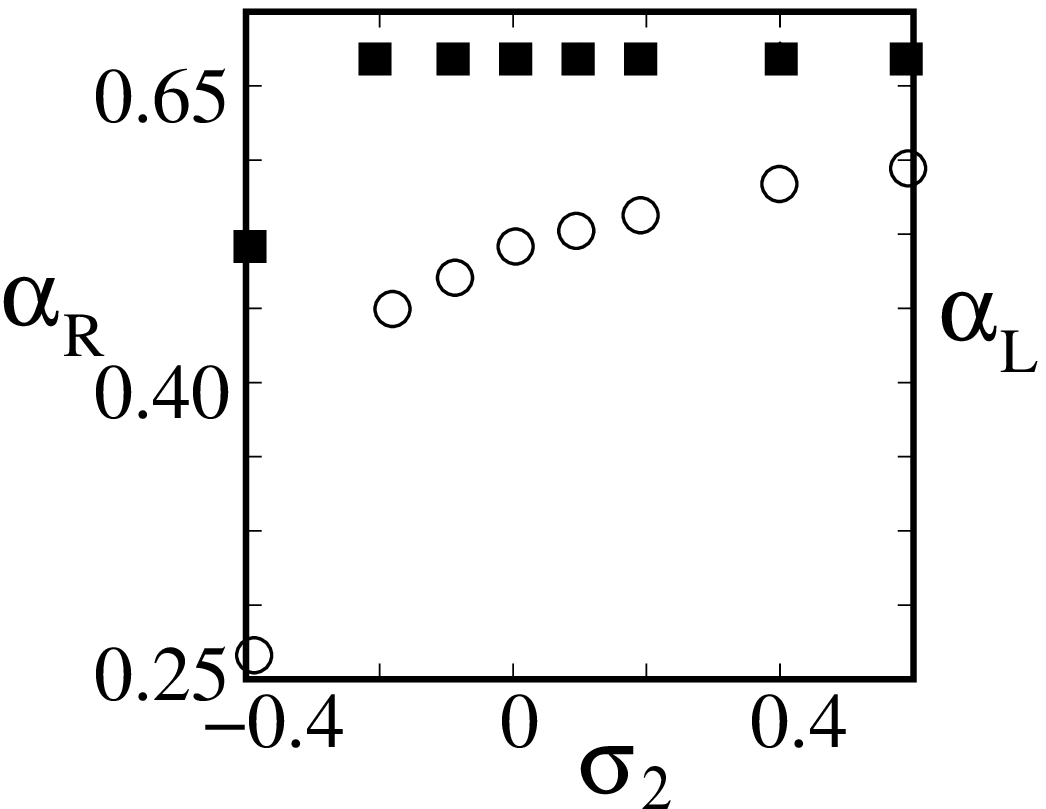}}{\vskip-.0in
Fig. \ 5~~  Dynamic exponents of $\alpha_R$ and $\alpha_{\cal L}$
as a function of  $\sigma_{-1}$  for $\sigma_1=1$.
The open circles represent $\alpha_{\cal L}$ and 
the dark squares represent $\alpha_{R}$ }
  
\myfigure{\epsfysize1.1in\epsfbox{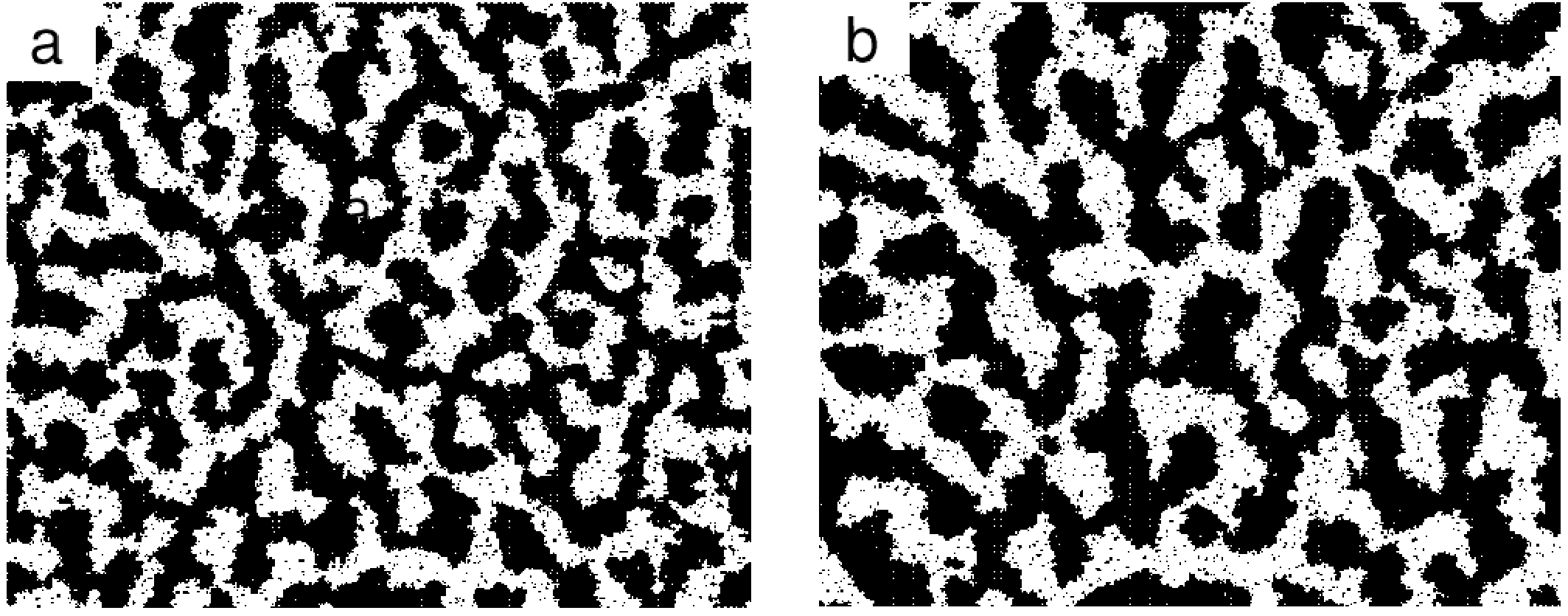}}{\vskip-.0in
Fig. \ 6~~  Time evolution of binary mixture for d=0.55, $\sigma_1= 1.0, 
\sigma_{-1}=-0.4$. The snap shots are taken at (a)$t=25118$ and (b)$t=39810$.  }

We now investigate the more general case of $\sigma_1 \ne \sigma_{-1}$
 with $\sigma_{-1}< \sigma_1$.
The dynamic exponents $\alpha_R$ and $\alpha_{\cal L}$ characterising the 
time dependence of 
$R$ and ${\cal L}$ seem to vary differently with  $\sigma_1-\sigma_{-1}$
This is depicted in figure 5. This results imply break down of scaling .
The absence of scaling is clearly manifested in the snap shots 
given in figure 6, the coarsening process
 is through the breaking and joining of domains.
Since the length $R(t)$ is decided by the big clusters it has a
dynamical exponent characteristic of coalescence. On the other hand 
the total border length $ {\cal L}(t)$ has significant contribution
from the small clusters which coarsen through evaporation condensation 
resulting in an exponent $1/3$.

\myfigure{\epsfysize1.1in\epsfbox{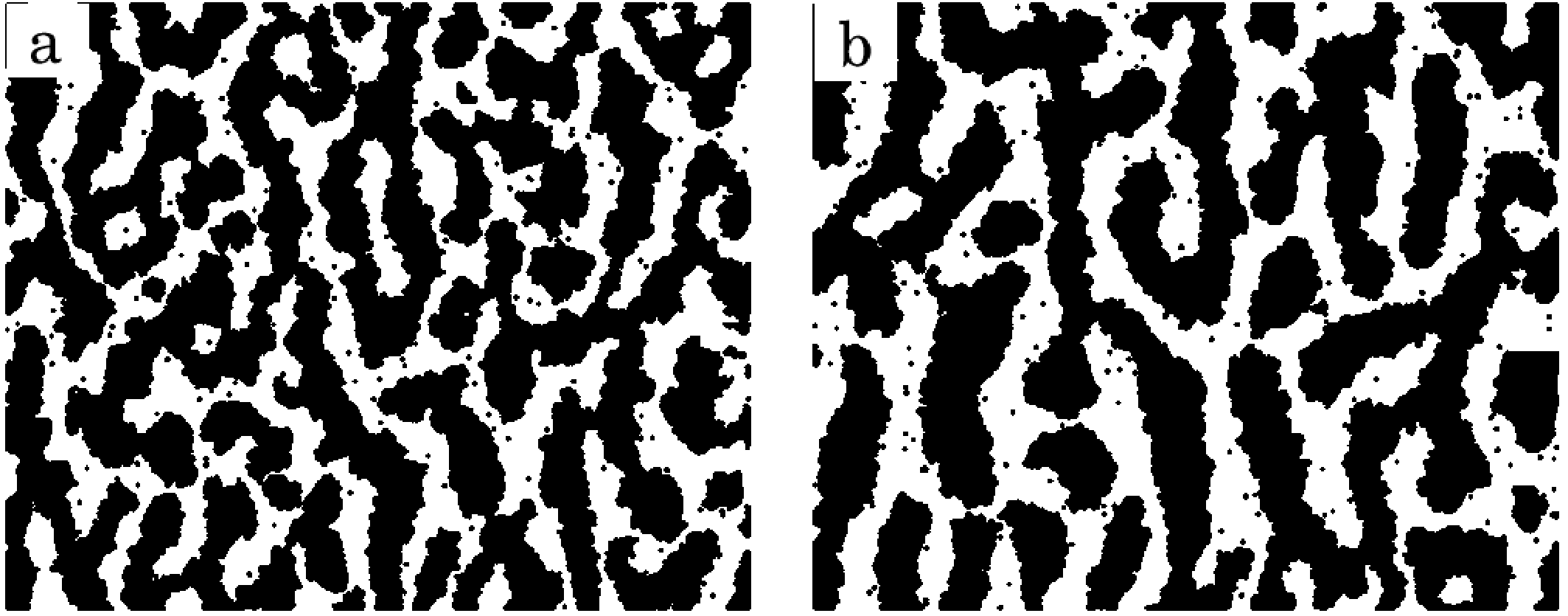}}{\vskip-.0in
Fig. \ 7~~  Time evolution of binary mixture in the RK model
 for d=0.55, $\sigma= 1.0, \beta=0.03$. The snap shots are taken 
at (a)$t=25118$ and (b)$ t=39810$.  }

\myfigure{\epsfysize1.8in\epsfbox{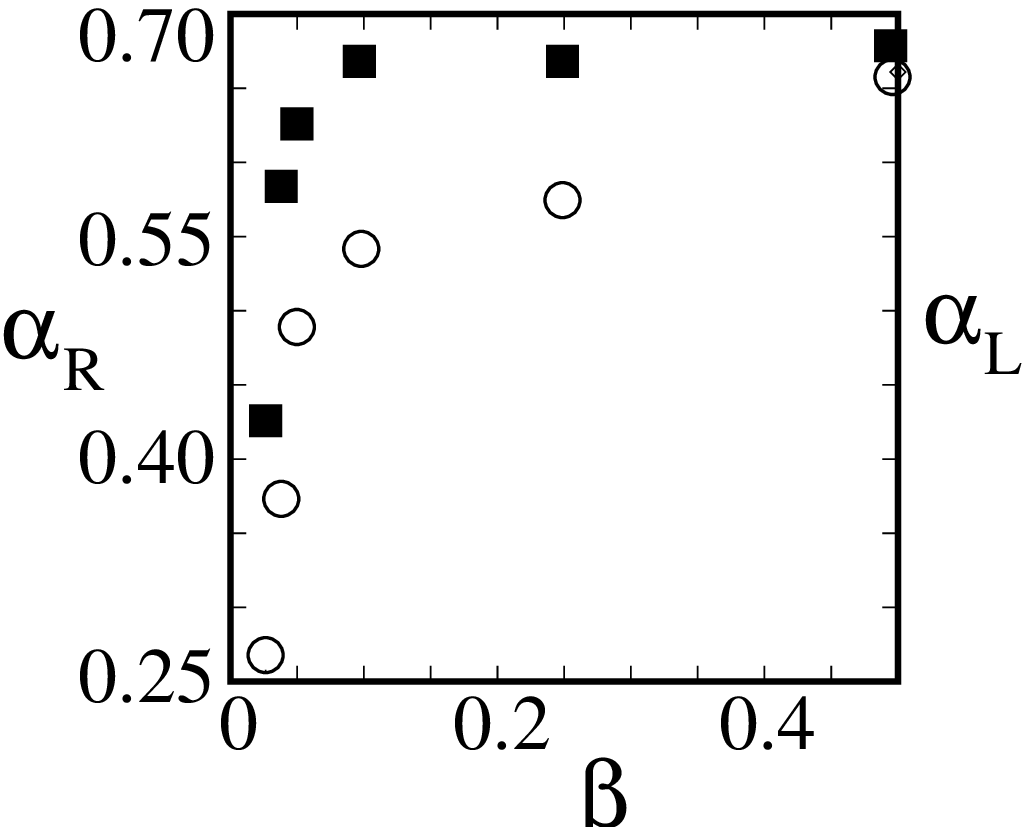}}{\vskip-.0in
Fig. \ 8~~  Dynamic exponents of $\alpha_R$ and $\alpha_{\cal L}$
as a function of  $\beta$  for $\sigma=1$ in the RK model.
The open circles represent $\alpha_{\cal L}$ and 
the dark squares represent $\alpha_{R}$ }

Similar change in the dynamical exponents of $R$ was observed as a 
function of the inverse temperature like parameter $\beta$ in the 
RK model ~\cite{coveney-emerton97}. 
 For low $\beta$ values the coarsening 
mechanism very similar to that operating in the case of 
$\sigma_1\ne\sigma_{-1}$in the GR model. This is shown in figure 7. 
The time dependence of $R$ and ${\cal L}$ shown in figure 8 confirms 
absence of scaling. These exponents indicate  
the presence of more that one coarsening mechanism operating at the 
late time regime. 

In the case of two fluid mixtures with unequal volume fraction  we see 
the initial LSW regime crossing over to the $\alpha=1/2$ regime controlled 
by the droplet coalescence mechanism. The first zero of the correlation function
$R$ and the 
total border length ${\cal L}$ in this regime seems to indicate existence 
of simple scaling ~\cite{wagner-cates01}.

\section{Three Component Fluids}

In this section, we look at two separate cases of phase separating 
three component fluids.In one, we consider a {\em symmetric ternary mixture}
with all the components having equal volume fraction and 
with the parameters $\sigma_i$ in equation-\ref{three-energy}  
set to $\sigma_1=\sigma_{-1}=\sigma_0$.  
This amounts to having the same surface tension 
for the $A-B, B-C $ and $A-C$ interfaces.  The other case of interest 
 is the {\em asymmetric mixture}, where we choose the parameters 
to be $\sigma_1=\sigma_{-1}$ and $\sigma_0<2(\sigma_1+\sigma_{-1})$. This 
choice of parameters ensures that one of the components ({\em solute}) is 
equally soluble in the other two. We will now discuss these two cases in detail.

\subsubsection {Symmetric case}
\myfigure{\epsfysize1.6in\epsfbox{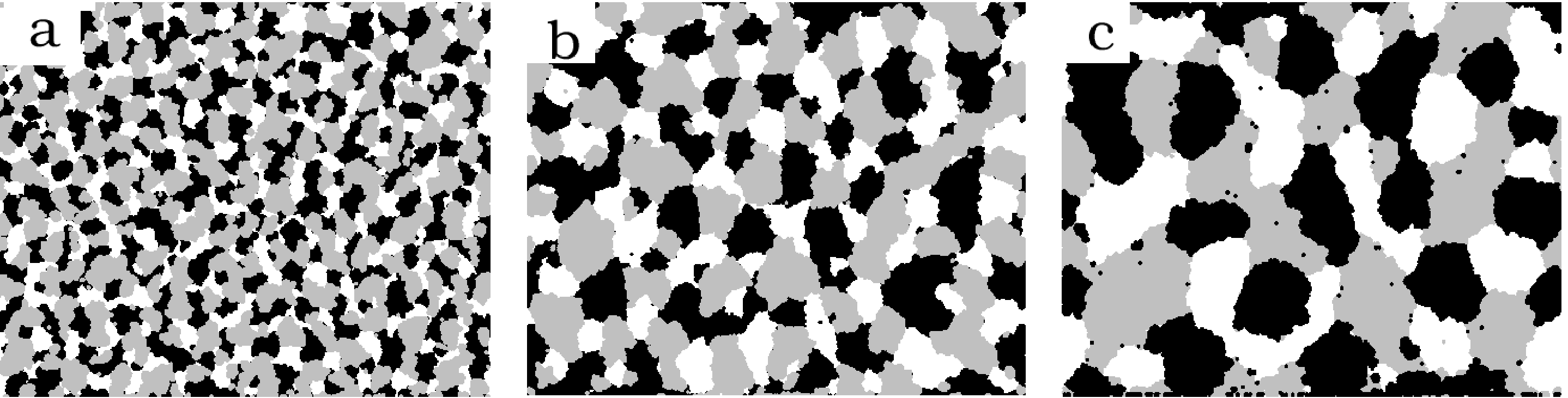}}{\vskip-.0in
Fig. \ 9~~  Time evolution of three fluid mixture for d=0.5, $\sigma_1=\sigma_{-1}=\sigma_0=1$. The snap shots are taken at (a)$t=10000$,(b)$ t=39810$,(c)$ t=100000$.  }

Starting from the mixed state, the symmetric ternary mixture, when quenched below 
the transition temperature,  evolves to 
form droplets of individual components.  Snap shots from the simulation,
shown in figure 9,
clearly establish the existance of  sharp interfaces between the components. 
The first zero of correlation  function 
$R$ and the interface length ${\cal L}$ are plotted in figure 10 as a function 
of time.

Initial regime corresponds to the formation of droplets.
In figure-10 we see a region 
with $R(t) \sim t^{1/2}$, 
where this droplets coalesce to form bigger domains. If the late time growth 
is due to the coalescence  by droplet diffusion, as it is believed to be, then 
the following simple argument shows that the exponent should be $1/2$.

\myfigure{\epsfysize1.8in\epsfbox{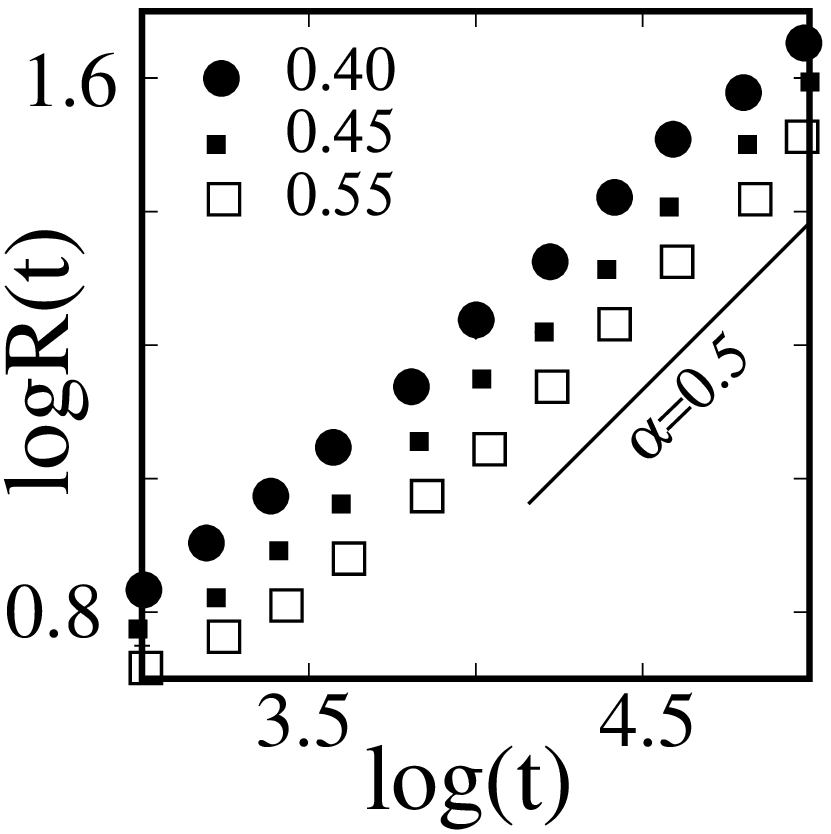}}{\vskip-.0in
Fig. \ 10~~ Correlation  length as a function of time in the immiscible
three component case for densities are 0.45,0.5 and 0.55. $\alpha =.5$ line 
is given as a guide to the eye.  }

If R  is the typical radius of a droplet, we have the droplet number
density    $ n \sim v/R^{2}$, where $v$ is the volume fraction of the 
components in two dimensions.  The time for a droplet to diffuse a distance 
of order of its radius is $ t_{R} = R^{2}/D $ , where $D$ is the diffusion coefficient. The area swept out by the
drop in time t (for $ t > t_{R}$ )is of the order 
 $R^{2} t/t_{R} \sim Dt$ . If $ t_{c} $ is the coalescence time , the expected 
number of drops in an area $ Dt_c $  is of the order unity, which means $nDt_{c}=1$. Since 
in two dimensions, the diffusion coefficient,
$D\sim k_B T/\eta$, does not have any dependence on the droplet size,

\begin{equation}
  tc \sim \frac{R^{2}}{vD}  \sim \frac{\eta R^{2}}{vk_{B}T}.
\end{equation}

  This implies that  R grows with time as 
$ R \sim (\frac{vk_{B}Tt}{\eta})^{1/2}$ ~\cite{siggia73}.      

Another mechanism for droplet coalescence is interfacial diffusion. At non 
zero temperatures the domains fluctuate from their circular shape. The area
explored by the domain in time $t$ goes as $D_b t$, where $D_b$ is the 
diffusion coefficient for the interface.The coalescence time $t_c$ is then 
given by 

\begin{equation}
t_c\,\sim \, \frac{R_0^2}{D_b},
\end{equation}

here the length $R_0$ is such that $\pi R_0^2\,=\,$ area of the droplet. Since 
the interface diffusion coefficient is not a function of $R_0$, we get $L(t)\sim t^{1/2}$. Thus both the droplet diffusion and interfacial diffusion could give 
the same dynamical exponent. 

To explore the possibility of the second mechanism we studied the stability of 
a drop to fluctuations from the circular shape. As shown in the appendix,
in two dimensions, surface fluctuations die out exponentially, both in the case 
of a drop and a strip ~\cite{sanmiguel-grant85} of one fluid in another.
This implies that in the $\alpha=1/2$ regime, the dominant mechanism for 
coarsening is the droplet coalescence by diffusion.

\myfigure{\epsfysize1.8in\epsfbox{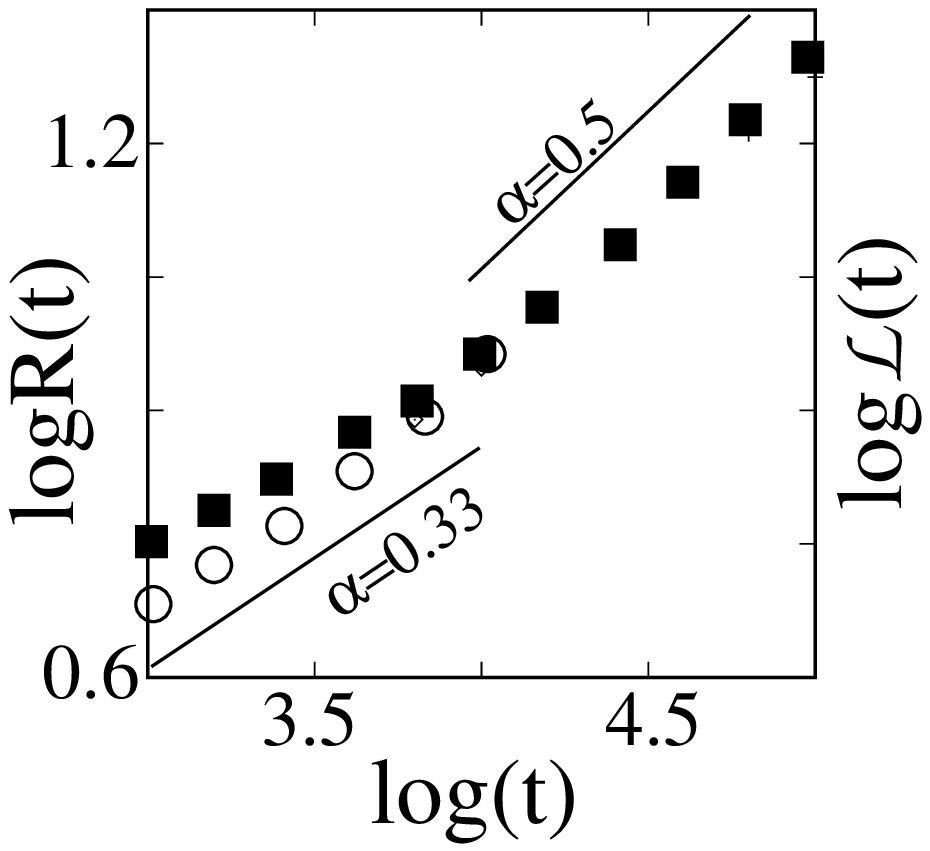}}{\vskip-.0in
Fig. \ 11~~ The first zero of the correlation function R(open circles) 
and the total border length ${\cal L}$ (filled squares) for a symmetric
ternary fluid with $\sigma_1 =\sigma_{-1} =\sigma_0$ as a function of time.
Average density of the fluids is d=0.55. 
Continuous lines with slopes 
$1/3$ and $1/2$ are given as guide to the eye}.

Droplet coalescence is the only mechanism operating here is further confirmed 
by the existence of scaling. In figure 11 we compare the time dependence of the
 total border length ${\cal L}_A$ of the domains of component A
with that of the 
first zero of the correlation function
$R_{11}(t)$.  We observe the same time dependence for both the lengths 
indicating scaling in the late time regime.

\subsubsection {Asymmetric case}
\myfigure{\epsfysize1.91in\epsfbox{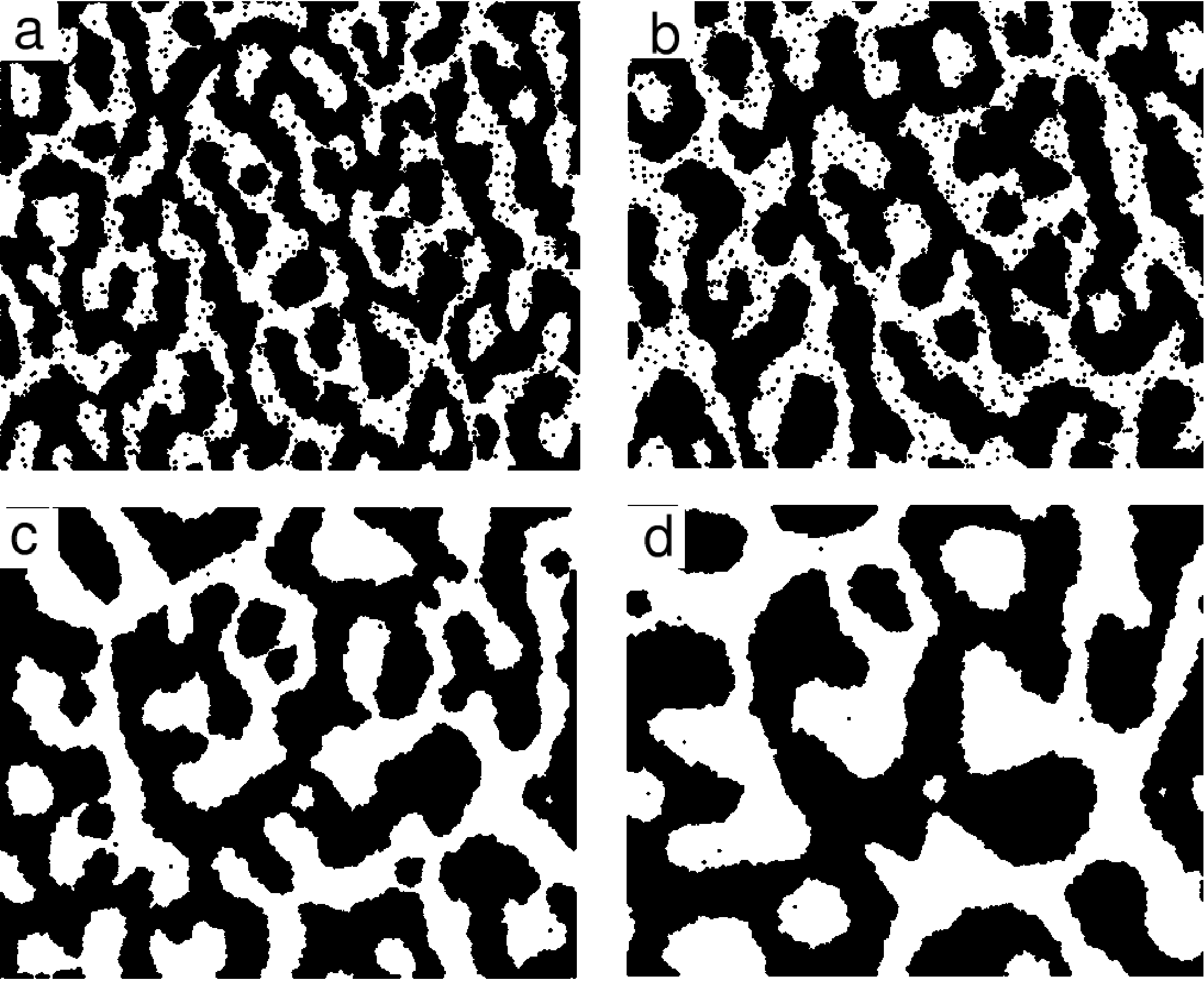}}{\vskip-.0in
Fig. \ 12~~  Time evolution of three fluid mixture for d=0.55 . (a) and (b) 
are the snap shots from the asymmetric ternary mixture phase separation at
$t=25118$ and $t=39810$ with $\sigma_1=\sigma_{-1}=1$ and $\sigma_0=-1.5$. Note the breaking and reorganization of the 
domains with time. A-C region is shown in black and B-C region in white.
  For comparison we show snap shots (c) and (d) from the phase 
separation of a binary mixture having the same density at the same time steps with $\sigma_1=\sigma_{-1}=1$.}

In this section, we discuss a ternary mixture with one of the components 
equally soluble in the other two. For this we choose 
$\sigma_1 = \sigma_{-1}$ and 
$\sigma_0 < -(\sigma_1+\sigma_{-1})$
such that the component C ($i=0$) is  the solute. The components A and B 
are mutually immiscible.  We choose A and B to have the same volume fraction.  
We now have A-C and B-C mixtures phase separating  similar to 
the binary fluid under critical quench. 

As shown in figure 12, 
the starting mixed phase separates into percolating domains 
of A-C and B-C phases. Though the early time behavior is similar to the two
component fluid with $\sigma_1=\sigma_{-1}$, 
the late time growth is analogous to the situation where
 $\sigma_1 \ne \sigma_{-1}$ discussed in section III. The snap 
shots do not exhibit self-similarity,  domains
split up and combine  during coarsening.
In this 
regime, the first zero of the correlation function exhibits a dynamic exponent $\alpha=1/2$.
We would like to point out that for the same density, the  
two component fluid with $\sigma_1=\sigma_{-1}$ exhibits a late
time exponent $\alpha =2/3$ and does not show 
any evidence of $\alpha=1/2$ regime.  Thus the third component 
act to reduce the interfacial tension between the $A$ rich and $B$ rich phases 
leading to fluctuation induced  breakup and reorganization of the domains.
As the volume fraction of 
the component C is reduced, we observe a gradual change from the 
$\alpha=1/2$ to $\alpha=2/3$ regime.

In order to investigate the validity of the scaling hypothesis in this 
regime, we compare the length scales ${\cal L}(t)$ and $R(t)$. This is 
depicted in figure 13. It is evident from the figure that these two length 
scales vary differently at late times.

\myfigure{\epsfysize1.8in\epsfbox{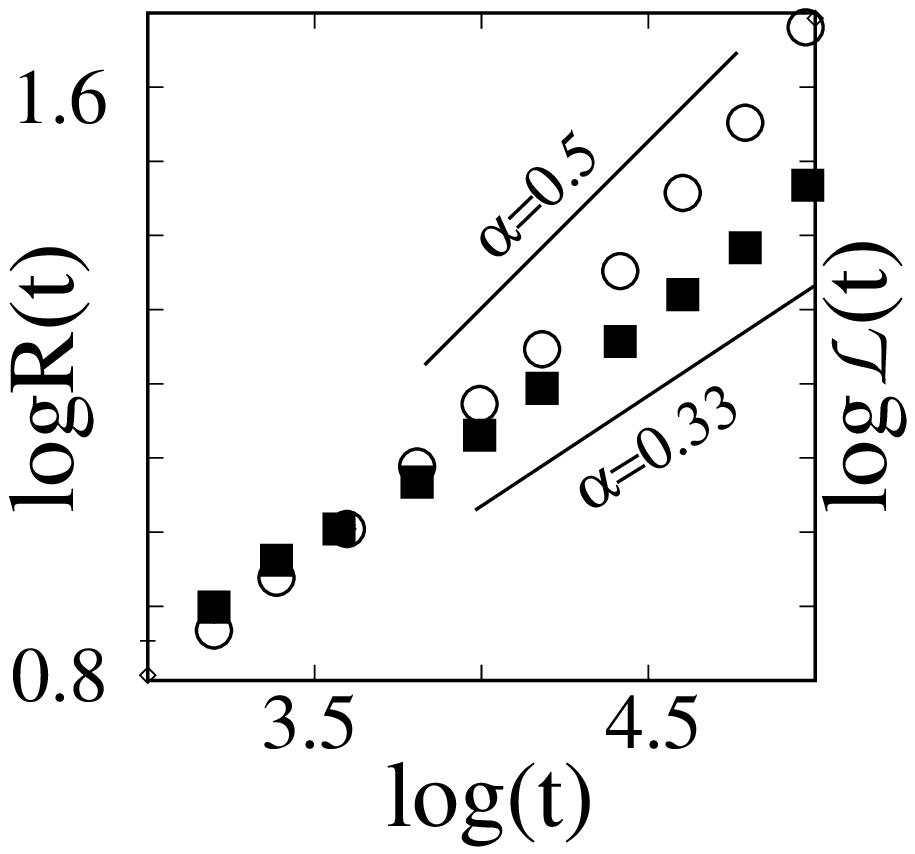}}{\vskip-.0in
Fig. \ 13~~ The first zero of the correlation function (open circles) and 
the total border 
length (filled squares) in three component asymmetric fluids.
Continuous lines with slopes  $0.33$ and $0.5$ are given as reference} 

\section{Conclusion}

We have presented the dynamics of domain growth 
in binary and ternary immiscible fluids ,
using a two-dimensional hydrodynamic lattice gas model. 
Various dynamic regimes are investigated by altering  
parameters like volume fraction, the interaction strength 
of the individual components and a temperature variable.  

We examine the validity of scaling in two component fluid coarsening
by comparing the dynamic nature of the 
first zero of the correlation function and the total border length.
While scaling holds in the droplet coalescence and the inertial 
hydrodynamic regimes, violation of scaling is observed in the fluctuation 
dominated regime. 

In symmetric ternary mixtures,
we show the existence of a coalescence dominated regime with a single 
length scale exhibiting a dynamical exponent $\alpha=1/2$.
In asymmetric ternary mixtures we use the concentration of the solute
as a control parameter. We demonstrate that at higher solute concentrations,
the coarsening is driven by fluctuations and scaling is violated.

\section{appendix}

\subsection{Stability of the drop}

In this appendix, we look at the stability of a drop with radius $R$ 
against perturbations in its perimeter.
We consider a drop of radius $R$ with the  fluid in the region 
$r<R$ characterized by density $\rho$, and kinematic viscosity 
$\nu=\frac{\eta}{\rho}$ 
surrounded by another fluid with density $\rho^{'}$ and kinematic viscosity 
$\nu^{'}$.
For an incompressible and  vorticity free fluid  we can define the velocity 
component using a  streaming potential $\psi$ ss 
$ v_r= \frac{1}{r} \frac{\partial{\psi}}{\partial \phi} $
$ v_{\phi}= -\frac{\partial{\psi}}{\partial r} $
	
Navier-Stokes equation then reduces to  

\begin{equation}
   (\partial_{t}-\nu \nabla^{2})\nabla^{2}\psi (r,\phi,t) =0
\end{equation}
	
We look for solutions of the form,$
 \psi(r,\phi,t) = \psi_{1}(r,\phi,t) + \psi_{2}(r,\phi,t) $
such that  
\begin{eqnarray}
 \nabla^{2} \psi_{1}&=&0. \\
   (\partial_{t}-\nu \nabla^{2})\nabla^{2}\psi_{2}(r,\phi,t) &=&0.
\end{eqnarray}

The general solutions of these equations in cylindrical polar coordinates 
 is 
\begin{equation}
\psi_{1}(r,\phi,t) =(a_{m} r^{m} + \frac{b_{m}}{r^{m}})  e^{im\phi} e^{\omega t}.
\end{equation}
\begin{equation}
\psi_{2}(r,\phi,t)=(\alpha_{m} I_{m}(kr)+ \beta_{m} K_{m}(kr)) e^{im\phi} 
                    e^{\omega t}.
\end{equation}
	Where $k=(\omega/\nu)^{1/2}$ and $I_{m}(kr)$ and $K_{m}(kr)$ are 
spherical Bessel functions.

\vspace{1.0cm}

The coefficients $a_m,b_m,\alpha_m$  and $\beta_m$ are obtained from the 
four equations; the continuity of $v_{r}$ at $r=R$ ,
continuity of $v_{\phi}$ at $r=R$, continuity of tangential stress 
$\sigma_{r\phi}$ at $r=R$ and the relation between normal stresses 

\begin{equation}
\sigma_{rr}|_{r=R^-} +p^- =\sigma_{rr}|_{r=R^+} + p^+ + p_\sigma.
\end{equation}

For the pressure $p$ we write $P=p(r)e^{im\phi} e^{\omega t}$. The pressure 
due to surface tension is obtained from the Young Laplace formula 
\begin{equation}
 P_{\sigma} =\frac{\sigma i m^{3}}{R_{0}^{3} \omega} \psi_{r<R}.
\end{equation}.

These equations can be written in a matrix form, 

\begin{equation}
\left[ \begin{array}{cccc}
R^m  &-1/R^m &  I_m  & -K_m \\
R^{m-1} &1/R^{m+1}& -I_{m}/R +\frac{k}{m} I_{m-1}& -K_{m}/R +\frac{k}{m} K_{m-1}\\
-2(m^{2}+m)R^{m-2} & {\frac{2(m^{2}+m)}{R^{m+2}}} &A_{33} & A_{34}\\
A_{41} &A_{42} &A_{43} & A_{44}
\end{array} \right] \left[ \begin{array}{c} 
a_m\\b_m\\ \alpha_m\\ \beta_m \end{array} \right] \,=\,0
\end{equation}

where, 
\begin{eqnarray}
A_{33}&=&-(\frac{k^{2}}{2} +\frac{2m}{R^{2}} +\frac{m^{2}}{R^{2}}) I_{m} 
	+\frac{mk}{2R} I_{m+1} 	+(\frac{mk}{2R} + \frac{k}{R}) I_{m-1} 
         - k^{2} I_{m-2} \\ \nonumber
A_{34}&=&(\frac{k^{2}}{2} +\frac{2m}{R^{2}} +\frac{m^{2}}{R^{2}}) K_{m} 
	+\frac{mk}{2R} K_{m+1} 	+(\frac{mk}{2R} + \frac{k}{R}) K_{m-1} 
         + k^{2} K_{m-2} \\ \nonumber
A_{41}&=& \nu( 2m^{2}+k^{2}R^{2}-1 +\frac{\sigma m^{3}}{R\omega \eta})R^{m-2}\\ \nonumber
A_{42}&=&\frac{\nu( 2m^{2} +k^{2}R^{2} -1)}{ R^{(m+2)}}\\ \nonumber
A_{43}&=&\nu((\frac{2m^{2}}{R^{2}} -\frac{4m}{R^{2}}+\frac{1}{R^{2}} 
       +\frac{\sigma m^{3}}{R^3 \omega \eta})I_{m} - \frac{k}{mR}(1-2m) I_{m-1}
       +\frac{2k}{R}(m-1) I_{m-1})\\ \nonumber
A_{44}&=&\nu((-\frac{2m^{2}}{R^{2}} +\frac{4m}{R^{2}}-\frac{1}{R^{2}})K_{m} 
        - \frac{k}{mR}(1-2m) K_{m-1} + \frac{2k}{R}(m-1) K_{m+1})\\ \nonumber
\end{eqnarray}

In the limit of large viscosity, we can neglect all terms of order 
$\sqrt(\omega/\eta)$.  The condition for non trivial solutions 
 in this 
limit leads to a relation between $\omega$ and $m$;
\begin{equation}
   \omega =\frac{m^{3} \sigma}{(1-4m^2) R \eta} . 
\end{equation}

Since the right hand side of this equation is always negative, we come to 
the conclusion that  in the limit of high viscosity there are no unstable 
modes. In the more general case, the condition for non trivial solutions 
leads to a transcendental equation for $\omega$ and $m$. A numerical 
analysis on this equation does not yield any unstable solutions.


\begin{thebibliography}{99}
 
\bibitem{wong-knobler81}Wong N C and Knobler C M 1981 {Phys. Rev. A},{\bf 24},
3205

\bibitem{chou-goldburg81} Chou Y C and Goldburg W I 1981 {Phys. Rev. A},{\bf 23},
858
 
\bibitem{perrot-guenoun94} F.Perrot,D.Guenoun,T.Baumberger and D.Beysens,
 {Phys. Rev. Lett},{\bf 73},688 (1994)

\bibitem{walheim99} S. Walheim, M. Ramstein, and U. Steiner., Langmuir., {\bf 15}, 4828(1999)

\bibitem{velasco-toxvaerd93} E. Valasco and s. Toxvaerd , {Phys. Rev. Lett.},{\bf71}
,388 (1993)
 
\bibitem{leptoukh-strickland95} G.Leptoukh,B.Strickland and C.Roland , {Phys. Rev. Lett.},{\bf 74} ,3636 (1995)

\bibitem{valls-farrel89} O.T.Valls and James.E.Farrell,
{Phys. Rev. B},{\bf 40} ,7027-7039(1989)

\bibitem{bray94} A.J.Bray  ,  {Adv. Phys},{\bf 43} ,557 (1994)

\bibitem{siggia73} E.D.Siggia , {Phys. Rev. A},{\bf 20} ,595 (1973)
\bibitem{furukawa94} H.Furukawa  , {Physica A},{\bf 204} ,237 (1994)
\bibitem{sanmiguel-grant85} M.San Miguel,M.Grant and J.D.Gunton  ,
{Phys. Rev. A},{\bf 31} ,1001-1005 (1985)


\bibitem{yeoman1-96} M.R.Swift,E.Orlandini,W.R.Osborn,J.M.Yeomans,  {Phys.Rev. E},
{\bf 54},5041-5052 (1996)
 
\bibitem{chen93} F.J.Alexander,S.Chen, and D.W.Grunau  {Phys.Rev. B},
{\bf 48},634-637 (1993)
 
 
\bibitem{yeoman3-98} A.J.Wagner and J.M.Yeomans , {International journal of
Modern physics},{\bf 9},1373-1382  (1998)           

\bibitem{wagner-yeomans98} A.J Wagner and J.M.Yeomans, Phys. Rev. Lett. 
{\bf 80},1429 (1998)

\bibitem{tanaka94} H.Tanaka. Phys. Rev. Lett. {\bf 72} ,1702(1994)

\bibitem{furukawa00} H. Furukawa, Phys. Rev. E. {\bf 61},1423(2000) 

\bibitem{wagner-cates01}  A. J. Wagner and M. E. Cates.
Euro. Phys. Lett., {\bf 56}, 556 (2001)

\bibitem{pagonabarraga01} I. Pagonobaraga, A. J. Wagner and M. E. Cates.
{\it To appear in } J. Stat. Phys. (2002)

 \bibitem{rothman-keller88} D.H.Rothmann and J.M.Keller , {J.Stat.
Phys.} ,{\bf2} ,1119-1127 (1988)

 \bibitem{bastea-lebowitz95} S.Bastea and J.L.Lebowitz ,{Phys.Rev.E} ,
{\bf52} ,3821-3826(1995)

 \bibitem{coveney-emerton97} A.N.Emerton,P.V.Coveney, and B.M.Bhogosian
  Phys. Rev. E ,{\bf55} ,708(1997)


 \bibitem{kawakatsu93} T.Kawakatsu {\em et. al} J. Chem. Phys ,{\bf99} ,8200(1993)

 \bibitem{Gonnella98} G. Gonnella, E. Orandini and J. M. Yeomans {\em et. al}
Phys. Rev. E,  {\bf58} ,8200(480)


\bibitem{laradji-mouritsen94} M.Laradji,O.G.Mouritsen and S.Toxvaerd  ,
{Europhys. Lett.},{\bf 28} ,157-162 (1994)  

\bibitem{smith00} K. A. Smith {\it at. al} Phys. Rev. Lett. {\bf 84}, 91 (2000)
             
 \bibitem{gunstensen-rothman90} A.K.Gunstensen and D.H.Rothman , {Physica D} ,
{\bf47} ,47-52(1990)

 \bibitem{chopard-droz98} B.Chopard and M.Droz , {Cellular automata modeling of physical systems},
     Cambridge University Press (1998)
 
                              
\end{thebibliography}
\end{document}